\documentclass[prb,superscriptaddress,twocolumn,amsmath,amssymb,floatfix]{revtex4-1}

\usepackage{latexsym}
\usepackage{dcolumn}
\usepackage[dvips]{graphicx}
\usepackage{amssymb}
\usepackage{graphicx}
\usepackage{psfrag, color}
\usepackage{verbatim}
\usepackage{amsmath}
\usepackage{epsf}

\begin{document}
\draft

\title{Electronic transport in two dimensional Si:P $\delta$-doped layers}
\author{E. H. Hwang}
\affiliation{Condensed Matter Theory Center, Department of Physics, 
	 University of Maryland,
	 College Park, Maryland  20742-4111}
\affiliation{SKKU Advanced Institute of Nanotechnology, Sungkyunkwan
  University, Suwon 440-746, Korea } 
\author{S. Das Sarma}
\affiliation{Condensed Matter Theory Center, Department of Physics, 
	 University of Maryland,
	 College Park, Maryland  20742-4111}

\date{\today}

\begin{abstract}
We investigate theoretically 2D electronic
transport in Si:P $\delta$-doped layers limited by charged-dopant
scattering. Since the carrier density is approximately equal to the dopant
impurity density,
the density dependent transport shows qualitatively different behavior from
that of the well-studied 2D
Si-MOSFETs where the carrier density is independent of the impurity
density. We find that the density dependent
mobility of the Si:P system shows non-monotonic behavior
which is exactly opposite of the non-monotonicity observed in
Si-MOSFETs --- in the Si:P system the mobility first decreases with
increasing density and then it increases slowly with increasing
density above a typical density $10^{14}$ cm$^{-2}$ (in contrast to
Si MOSFETs where the mobility typically increases with density first
and then slowly decreases at high density as surface roughness
scattering dominates). In the
low density limit (or strong screening limit) mobility
decreases inversely with increasing density, but in the high density
limit (or weak screening limit) it slowly increases due to the finite width
effects of the 2D layer. In the intermediate density regime ($1/a < 2 k_F <
q_{TF}$, where $a$, $k_F$, and $q_{TF}$ are the confinement width of
the $\delta$-layer, Fermi wave vector, and Thomas-Fermi screening wave vector,
respectively) the density 
dependent mobility is approximately a constant at the minimum value. 
However, the calculated mean free path
increase monotonically with density.
We also compare the transport scattering time
relevant to the mobility and
the single particle relaxation time relevant the quantum level
broadening, finding that the transport scattering time could be much
larger than the single-particle scattering time unlike in Si MOSFETs
where they are approximately equal. 
\end{abstract}

\maketitle

\section{introduction}

In semiconductor based quantum computers, where electron spin (or
charge) localized at dopant sites could be used as
qubits\cite{kane1998}, it is extremely important to very precisely
place dopants with atomic precision within the host
lattice\cite{koiller2001}.  This has motivated a materials science
initiative to create precisely located dopant structures within
semiconductor hosts, e.g. P in Si or Ge.  In particular, the scanning 
tunneling microscopy (STM) based lithography\cite{lyding1994} 
has been used to position dopants with atomic scale
precision \cite{schofield2003} within semiconductors.
The ability of controlling the location of individual dopant atoms within
a semiconductor provides future potential towards atomic scale
devices\cite{kane1998,koiller2001,snider1999}. Recently,
Si:P $\delta$-doped two dimensional (2D) structures in silicon have attracted 
interest with their potential applications in multiqubit quantum
computer architectures\cite{koiller2001} and atomic-scale device
\cite{shen2004,ruess2005,wilson2004,zudov2003,goh2006}.
The effect of doping density on electronic transport in Si:P
$\delta$-doped layers grown by phosphine dosing and low temperature
molecular beam epitaxy have been investigated in
experiments\cite{shen2004,ruess2005,wilson2004,zudov2003,goh2006,weber2012}. 
Unlike other 2D systems, the carrier concentration in this $\delta$-doped
layers can reach $10^{14}$ cm$^{-2}$ or higher, which 
exceeds a typical carrier density found in GaAs-based systems or
Si-MOSFETs by more than two orders of magnitude. 

Even though the carrier density of Si:P $\delta$-doped layers is
extremely high  
the measured mobilities\cite{zudov2003,goh2006,weber2012,fuechsle2010} at $4K$
are very low $\mu \sim 30-200$ 
cm$^2$/Vs in P doping levels of $\sim 10^{14}$ cm$^{-2}$. Thus, in these $\delta$ layers
the mean free path $l$ is of the order of only a few nm and the
corresponding transport relaxation time $\tau$ is in the femtosecond range. 
In addition, as the carrier density increases mobility decreases at
first, but the mean free path increases linearly. These behaviors are  
opposite to those found in extensively studied 2D Si MOSFET and
GaAs/AlGaAs systems.  
It is, of course, expected that the electronic transport mobility of the
Si:P $\delta$-doped systems is lower than that in
modulation-doped GaAs systems because the charged dopant impurities in
Si:P layers are
located inside the layer, but the mobility for the Si:P system is
substantially lower than Si MOSFET mobility also. However, the
density dependence of low-temperature 
transport in Si:P layers is unusual
because it is dominated entirely by the dopant
charged Coulomb
impurity scattering \cite{ando1982,dassarma2011}. The mobility increases with
increasing density 
when the charged impurity scattering dominates in
the high density regime. In Si-MOSFETs, due to 
the surface roughness scattering at high densities, the mobility
decreases with increasing carrier density in the high density regime
dominated by surface roughness scattering 
which however plays no role in the transport properties of Si:P
delta-layers since the carriers are located away from the surface
inside the semiconductor. 
It is also reported that in Si:P $\delta$-doped samples with the doping
density of $ \alt 10^{13}$
cm$^{-2}$ a non-Ohmic behavior is observed, suggesting that a transition to
Anderson localization for Si:P $\delta$-doped layers occurs below $n\sim
10^{13}$ cm$^{-2}$. The similar low-density localization behavior is also observed
in other $\delta$-doped 
samples like Si:Sb and Si:B $\delta$-doped layers \cite{agan2001}.

These experimental features of $\delta$-doped layers provide a unique
opportunity to  
study the carrier density dependent transport of ultrahigh density
2D systems limited by only charged Coulomb disorder.
Since the carriers remain confined to a dopant plane by
strong-confining electrical
fields, charged Coulomb scattering by the ionized dopants themselves
significantly limits the mobility.
In spite of there being very interesting experimental studies in the
ultrahigh density $\delta$-doped systems, 
the systematic theoretical investigation of transport properties has
not yet been carried out.  
The purpose of this paper is to theoretically study the transport properties
of Si:P $\delta$-doped layers.
It is obviously important to understand the detailed transport properties for
eventual applications to atomic-scale devices and semiconductor
multiqubit architectures. 

In addition to atomic scale device application motivation there are
several fundamental reasons which make transport studies of 2D
$\delta$-doped structures interesting and important in its own right.
First, these systems allow the study of 2D carrier transport
properties at unprecedented high carrier densities of $10^{14}$ cm$^{-2}$ or
above, which is much higher than the typical $10^{10}-10^{12}$ cm$^{-2}$ density
regime studied extensively in the context of GaAs heterostructures and
Si MOSFETs.  Second, these systems allow for studying carrier
transport limited almost entirely by long-range Coulomb disorder
(arising from the ionized dopants producing the carriers in the
system) without complications from other mechanisms often operational
in 2D semiconductor systems such as surface roughness scattering and
alloy disorder scattering.  Third,  the $\delta$-doped 2D systems enable
the study of carrier transport in the scattering regime where the
carrier density is roughly equal to (or perhaps even somewhat less
than) the impurity density since each electron must leave behind an
ionized impurity.  Fourth, by changing carrier density, one could
study the strong screening to the weak screening 2D transport regime,
which is not possible in other 2D systems.  All of these features are
unique to the $\delta$-doped 2D systems making them complementary to the
traditional 2D semiconductor systems and thus providing particular
fundamental impetus to our work.  We mention that 2D electron systems
possess the highly peculiar behavior, arising from the constant 2D
density of states which is energy independent. The low-carrier
density regime is the strong-screening regime (where $q_{TF} > k_F$) and the
high-density regime is the weak-screening regime (where $q_{TF} < k_F$), and
therefore, the ability to go to very high carrier density provides a
completely new perspective in the study of 2D semiconductor systems,
which is not available in the well-known Si MOSFET and GaAs
heterostructure based 2D systems. 

In this paper we investigate the low-temperature electronic transport in Si:P
$\delta$-doped ultra-high density 2D electron systems.
Employing Boltzmann transport theory, we
calculate density dependent mobility and scattering times of Si:P $\delta$-doped
layers in a variety of experimental situations. The electrons in
realistic $\delta$-doped systems,
where there are likely sources of additional scattering arising from
unintentional background impurities, 
may localize at low densities ($n \alt 10^{13} cm^{-2}$), but
in this paper we neglect localization effects and 
show our calculated results in the wide range of density
$ 10^{11} < n < 10^{15}$ cm$^{-2}$.
In fact, our theoretical results based on the semiclassical Boltzmann
theory give a clue about the density range of strong localization in
the system, which we discuss later in the paper.  All our explicit
results (as presented in our figures) for mobility, mean free path,
and scattering time explicitly assume that the 2D carrier density is
precisely equal to the 2D quenched charged impurity density, and
therefore, our mobility results are an upper bound to the experimental
mobility since the real systems studied in the laboratory are likely
to have additional impurities not related to the dopants used in
creating the electron gas.  We expect the measured mobility and
conductivity to be somewhat smaller than that calculated by us, but we
expect the experimental mobility to approach our theoretically
calculated results with improvement in the sample quality.  
The calculation within the density functional theory 
shows that the band structure of a Si:P $\delta$-doped layer
is a symmetric V-shaped 
potential well \cite{drumm2012,budi2012,lee2011,qian2005,carter2009}, 
but we use the square well model with a width $a$ as a
$\delta$-doped 2D layer. We confirm that our calculated 
results are not sensitive to the shape of the confinement
potential by also doing a calculation using a triangular confinement potential.
The calculated transport results depend only on the characteristic
confinement length '$a$' defining the thickness of the 2D electron
layer, and not on the details of the confining wavefunction itself.

The rest of this paper is organized as follows: in section II 
we provide the detailed transport theory and the analytic results for
both high and low densities, and in section III we show the 
numerical results of mobility and scattering times of Si:P
$\delta$-doped layers. 
We conclude in section IV with a discussion.

\section{theory}

The Si:P $\delta$-doped layer is a conducting layer of atomic
thickness, which is formed 
in a pure semiconducting crystal matrix due to dopant atoms located
within one crystallographic plane. The electrons in the potential
well form a 2D
electron gas in the plane of the $\delta$ layer, and behave as 
2D free carriers with appropriate
effective mass and spin/valley degeneracy quantum mechanically
confined by the ionized dopant-induced electric field within a
quasi-2D layer of thickness '$a$'.
The electrons in a Si:P $\delta$-doped layer are spatially confined along
the normal direction with respect to the layer of P dopants. In this
paper we designate the 2D layer as $(x-y)$ plane and the normal direction of
the layer as the $z$ direction ($z=0$ being the center of the well).
The wave function of an electron in the layer is given by
\begin{equation}
\psi({\bf r},z) = \exp(i {\bf k \cdot r}) \phi(z),
\end{equation}
where $\phi(z)$ is the confinement wave function in $z$-direction,
${\bf k}$ is the 2D 
electron wave vector in the plane, and ${\bf r} =(x,y)$. 
The calculated band
structure of Si:P $\delta$-doped layer depends sensitively on the
disorder model (i.e., the nature of symmetry, the exact placement of
dopants, other unknown fixed charges in the system,
and the size of the unit cell, etc.).
However, for simplicity, we assume
that the confinement profile is described by a square quantum well structure.
Although the quantum well may be a simplification of the actual situation
\cite{drumm2012,budi2012,lee2011}, 
our calculated results describe quantitatively the transport
properties of the Si:P $\delta$-doped layer
since the details of the confinement are not quantitatively important
except for the confinement width, which we parametrize.


To calculate the 2D mobility of Si:P $\delta$-doped layers,
we use the Drude-Boltzmann semiclassical
theory for 2D transport limited by the
scattering from charged impurities (i.e., dopants) \cite{ando1982,dassarma2011}.
We assume that the impurities (P dopants) are randomly distributed in
the x-y plane 
at $z=0$, i.e., the center of the quantum well and
the 2D carrier conductivity is entirely limited by charged impurity scattering.
Since the Fermi temperature is very high ($T_F \sim 700$ K) at the density
$n=10^{14}$ cm$^{-2}$ 
and the Bloch-Gr\"{u}neisen temperature at this density is very high
we neglect phonon scattering in this calculation. 
The Bloch-Gr\"{u}neisen temperature at the electron density $n=10^{14}$
cm$^{-2}$ becomes $T_{BG}=2k_F\hbar v_{\rm ph} \sim 250$K with the 
phonon velocity of Si $v_{ph} = 9.13 \times 10^5$ cm/s.
Below $T_{BG}$ acoustic phonons cease to be appreciably excited, and
no longer contribute to the relaxation rate \cite{kawamura1992}. 
We therefore include scattering by the random quenched charged dopants
as the only resistive mechanism in our theory. 

By considering the charged impurity centers located at $z=0$ with
an impurity density $n_i$ and
taking {\bf k} and ${\bf k'}$ to denote the 2D electron wave vector
before and after scattering, respectively, by a Coulomb scatterer (i.e.,
P dopant)  we have the transport scattering time 
in the Born approximation 
\begin{equation}
\frac{1}{\tau(\epsilon_{\bf k})} = \frac{2\pi n_i}{\hbar} \int 
\frac{d^2 k'}{(2\pi)^2} |{V_{i}({\bf q})}|^2
(1-\cos \theta_{\bf kk'}) \delta({\epsilon_{\bf k}-\epsilon_{\bf k'}}),
\label{eq:tau_in}
\end{equation}
where $V_{i}(q)$ is the screened Coulomb potential for 
electron-charged impurity interaction, ${\bf q = k-k'}$ is the momentum
transfer, $\theta_{\bf kk'}$ is the scattering angle, and $\epsilon_k =
(\hbar k)^2/2m$ is the electron energy.

The unscreened potential of a charged center located at $({\bf r}_i,z=0)$ is given by
$U_i({\bf r}) = \frac{e^2}{\kappa}\frac{1}{|{\bf r} - {\bf r}_i|}$,
where $\kappa$ is the background dielectric function of Si. The 2D
Fourier transform of $U_i({\bf r})$ becomes
$U_i(q) = v(q) F_i(q)$, where 
$v(q) = 2\pi e^2/\kappa q$ is the 2D bare Coulomb interaction and
$F_i(q)$ is the form factor of 
electron-impurity interaction and given by
\begin{equation}
F_i(q) = \int dz e^{-q|z|}|\phi(z)|^2.
\end{equation}
The form factor $F_i(q)$ becomes unity in the limit of vanishing
$q$. The screening effect can be included 
by dividing $U_i(q)$ by the RPA dielectric screening 
function, $\varepsilon(q)$, due to the 2D electrons themselves, 
\begin{equation}
\varepsilon(q) = 1+ v(q) F(q)\Pi(q),
\end{equation}
where $v(q)$ is the 2D bare Coulomb interaction, $F(q)$ is the form
factor for electron-electron interaction,
and $\Pi(q)$  is the 2D finite wave
vector polarizability function \cite{ando1982,dassarma2011}.
Thus, the screened Coulomb potential for electron-charged impurity
interaction becomes $V_i(q) = U_i(q)/\varepsilon(q)$.
The form factor $F(q)$ for electron-electron interaction associated
with the confinement wave function is defined by 
\begin{equation}
F(q)  =  \int dz \int dz' |\phi(z)|^2 e^{-q|z-z'|} |\phi(z')|^2,
\end{equation}
where $\phi(z)$ is a confining wave function in the $z$ direction.
When the $\delta$-doped impurities are located at $z$ from the center of the
confinement potential the electron-impurity Coulomb interaction has
the factor $e^{-q|z|}$ and becomes $V_i(q,z) = V_i(q) e^{-q|z|}$.

If we take the confining potential in the $z$ direction to be of
square-well form, then the quantized ground state wave function is given by
$\phi(z) = \sqrt{\frac{2}{a}}
\cos(\pi z/a)$ for $|z|<a/2$ and $\phi(z) = 0$ for $|z|>a/2$
with $a$ denoting the width of the square well.
In this case the form factors $F_i(q)$ and $F(q)$ are calculated as
\begin{equation}
F_i(q) = \frac{4}{qa} \frac{2\pi^2(1-e^{-qa/2}) + (qa)^2}{4\pi^2 +
  (qa)^2},
\end{equation}
and
\begin{equation}
F(q) = \frac{3(qa)+8\pi^2/(qa)}{(qa)^2+4\pi^2} -
\frac{32\pi^4[1-\exp(-qa)]}{(qa)^2[(qa)^2+4\pi^2]^2}.
\end{equation}

As a result of the external dc electric field ({\bf E}) driving the
electric current in the system the Fermi distribution function is 
deformed from equilibrium distribution $f_0(\epsilon_k) =
1/[e^{-\beta (\epsilon_k-\mu_0)} + 1]$ where $\beta = 1/k_BT$ is the
inverse temperature and $\mu_0$ is the chemical potential, i.e.,
$f(\epsilon_k)=f_0(\epsilon_k) + g(\epsilon_k)$ where $g(\epsilon_k)$ is 
proportional to the applied electric field. In the relaxation time
approximation \cite{ando1982} 
$g(\epsilon_k) = -\tau(\epsilon_k) e 
{\bf E \cdot v_k} \partial  
f_0(\epsilon_k)/ \partial \epsilon_k$, where ${\bf v_k}
= \hbar {\bf k}/m$ is the carrier velocity.
Thus, an induced current
density $j$ due to the electric field in a $\delta$-layer is given by
\begin{equation}
{\bf j} = ge^2 {\bf E} \int \frac{d^2k}{(2\pi)^2}{\bf
  v}_k^2 \tau(\epsilon_k) \left ( - \frac{ \partial  
f_0(\epsilon_k)} {\partial \epsilon_k} \right ),
\end{equation}
where $g$ is the total degeneracy including the usual spin degeneracy
of two plus any possible valley 
degeneracy arising from the bulk band structure of the system. 
From the definition of the drift mobility $\mu =
j/(en E)$ and the conductivity $\sigma = j/E$, we have
$\mu = {e} \langle \tau \rangle/m$
and $\sigma = {ne^2 \langle \tau \rangle}/m$,
where $m$ is the carrier effective mass and 
the energy averaged transport relaxation time $\langle \tau
\rangle$ is given by
\begin{equation}
\langle \tau \rangle = \frac {\int d\epsilon D(\epsilon) \epsilon
  \tau(\epsilon) \left [-\frac{\partial f_0(\epsilon)}{\partial 
  \epsilon} \right ]}  {\int d\epsilon D(\epsilon) \epsilon
   \left [-\frac{\partial f_0(\epsilon)}{\partial
  \epsilon} \right ]}.
\label{eq:tau}
\end{equation}
where $D(\epsilon) = g m/(2\pi \hbar^2)$ is the density of states with
total degeneracy $g=g_s g_v$, $\epsilon = (\hbar k)^2/2m$ is the 
2D electron energy dispersion,
and $\tau(\epsilon_k)$ is the energy dependent transport
relaxation time given in Eq.~(\ref{eq:tau_in}).

Since we are interested in the low-temperature density dependent
transport ($T\ll T_F$) of 
Si:P $\delta$-doped layers we first consider the zero temperature
limit. After performing the $k'$
integration in Eq.~(\ref{eq:tau_in}) and taking $k=k_F$ (i.e. the
Fermi wave vector) we have 
\begin{equation}
\frac{1}{\tau(k_F)} = \frac{1}{\tau_0} I(q_0),
\label{eq:taukf}
\end{equation}
where $q_0 = q_{TF}/2k_F$ ($q_{TF} =
g/a_B$ is a 2D Thomas-Fermi wave 
vector with effective Bohr radius $a_B = \hbar^2 \kappa /me^2$),
\begin{equation}
\frac{1}{\tau_0} =  {2\pi\hbar} \frac{n_{i}}{m}
(\frac{2}{g})^2q_0^2,
\end{equation}
and
\begin{equation}
I(q_0) =   \int_0^{1} \frac{dx}{\sqrt{1-x^2}} \left [ \frac{ x F_i(2k_F
    x)}{ x+ q_0 F(2k_F x)}
      \right ]^2.
\label{eq:iq}
\end{equation}
Due to the form factors it is not possible to get the full analytic
formula for Eq.~(\ref{eq:iq}). However, we can get asymptotic forms of
Eq.~(\ref{eq:iq}) for different density regions 
where large or small $q_0$ approximation may apply.
For Si(100) we have
$a_B \sim 30$\AA \; for electrons in the conduction band and $q_0 \approx 4.7
g^{3/2}/\sqrt{\tilde{n}}$, where $\tilde{n} = n/(10^{10}
cm^{-2})$. Thus, in the low density limit ($n < 10^{12} cm^{-2}$) $q_0 \ll
1$ (i.e., strong screening limit), and if $2k_F
a \ll 1$ (i.e. $a \ll 1/2k_F$), then the form factors are close to
unity and $I(q_0)$ is given by 
\begin{equation}
I(q_0)  \approx  \frac{\pi}{2q_0^2}.
\end{equation}
Now the scattering time becomes $\tau(k_F) \sim 2\tau_0 q_0^2/\pi
\propto 1/n_i$, which is independent of carrier density $n$. However in Si:P
$\delta$-doped layer the dopant
density $n_i$ is equal to the carrier density $n$ because every P atom
gives up one free electron to the layer
(as stated above, in real samples it is more likely that $n_i>n$ since
there could be background unintentional impurities in the system, but
$n=n_i$ provides a lower bound for the impurity density). 
Thus, the scattering time $\tau \propto 1/n$
and the density dependent mobility is inversely proportional to the density,
$\mu \propto 1/n$, in the strong screening limit ($q_0 \gg 1$).

In the high density limit ($n > 10^{14}$ cm$^{-2}$ which is the relevant density
range in the currently available Si:P $\delta$-doped layers) $q_0 \gg 1$ (i.e. weak
screening limit) we can approximately express Eq.~(\ref{eq:iq}) as
\begin{equation}
I(q_0) =   \int_0^{1} \frac{dx}{\sqrt{1-x^2}} \left [  F_i(2k_F
    x) \right ]^2 \;\;\;\; {\rm for} \;\; q_0 \ll 1.
\end{equation}
Then, we have
\begin{eqnarray}
I(q_0) & \approx & \pi/2  \;\;\;\;\;\;\; {\rm for}
\;\; 2k_F a < 1, \nonumber \\
    & \approx & {2}/{k_Fa} \;\;\; {\rm for} \;\; 2k_F a > 1.
\end{eqnarray}
Thus for $2k_F a < 1$ we have
$\tau(k_F) = \tau_0 \pi/2 \propto n/n_i$, and mobility $\mu \sim
n/n_i$. For $n_i = n$ the mobility is independent of carrier density.
For $2k_F a > 1$ we have
$\tau(k_F) = \tau_0 k_F a/2 \propto n^{3/2}/n_i$, and the mobility
becomes $\mu \sim n^{3/2}/n_i$. For $n_i = n$ the mobility
increases as $\sqrt{n}$ as the carrier density increases.

Combining all the results we have so far we can conclude 
that as the carrier
density (or doping density) increases the mobility decreases as $1/n$
up to $2k_F \sim q_{TF}$. As the carrier density increases further
beyond this point the
density dependent mobility saturates up to $2k_F \sim 1/a$, and then
it increases as $\sqrt{n}$ when $2k_F > 1/a$.
Thus the mobility behavior in the delta-doped layers as a function of
carrier density (or dopant density since we assume them to be the same
in our model) has three distinct regimes: a low-density
strong-screening regime where the mobility decreases (as $1/n$) with
increasing density, an intermediate-density saturation regime where
the mobility is approximately a constant in density, and then a
high-density weak-screening regime where the mobility rises with
density (as $n^{1/2}$).  This behavior, which is clearly manifested in our
full numerical results shown in the next section, is in sharp contrast
with 2D transport in Si MOSFETs where the mobility first increases
with carrier density as Coulomb disorder gets screened out and then
decreases with carrier density as surface roughness scattering becomes
dominant.  We note that if the system is strictly two-dimensional
(i.e. the electrons are confined in an infinitesimally thin 2D layer),
then the form-factors in Eq.~(\ref{eq:iq}) both become unity, and the mobility
will decrease monotonically with increasing carrier density finally
saturating at high enough density where $q_0 \ll 1$ condition is satisfied.
The third regime of an eventual $n^{1/2}$ increase of mobility at high
density arises entirely from the quasi-2D form factor effect which
becomes crucial at very high density when $k_F \gg 1/a$, which would not
apply if $a=0$, i.e. in the strict 2D limit. We emphasize again that our
theory is an upper bound to the expected experimental mobility since
we are only including scattering by  the charged dopants themselves
ignoring all other possible scattering sources.

\begin{figure}[t]
	\centering
	\includegraphics[width=1.\columnwidth]{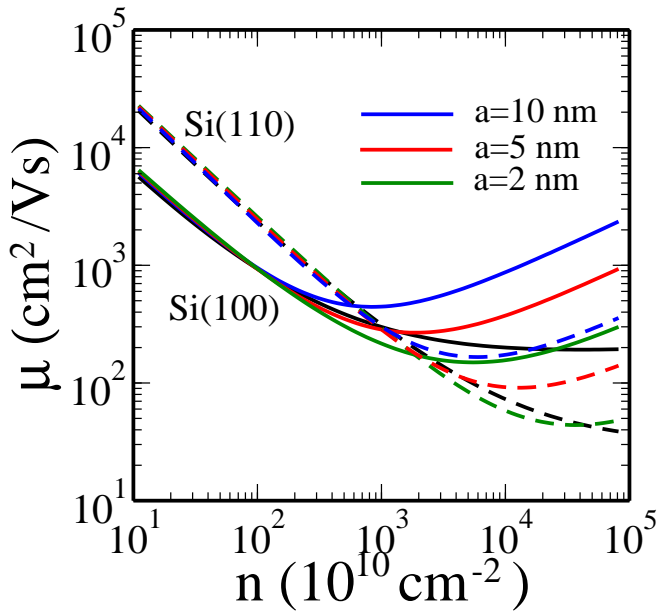}
	\caption{
Calculated mobility of Si:P $\delta$-doped layers as a function of
density at $T=0$ for different quantum 
well width $a=10$, 5, and 2nm. Black lines indicate the results
calculated with Howard-Fang (HF) variational wave function. Here
impurities are located at $z=0$ (i.e. the center of quantum well).
	\label{fig:fig1}}
\end{figure}

\section{results}

In this calculation we use the following parameters;
for Si(100) surface orientation the effective mass  
$m^*=0.19 m_e$ and valley degeneracy $g_v=2$, and 
for Si(110) surface orientation $m^*=0.4 m_e$ and $g_v=4$. 
We use the spin-degeneracy $g_s=2$ everywhere.
We set the carrier density to
be equal to the dopant impurity density, i.e., $n = n_i$.
We also take a square quantum well with width $a$ as a
confinement potential.
(We have explicitly verified that our numerical results change little
for a triangular potential well as long as the same effective quasi-2D
confinement width is used in the theory.)  We also mention that
mobility is simply related to the conductivity $\sigma$ by the formula
$\sigma=ne\mu$.

\begin{figure}[t]
	\centering
	\includegraphics[width=\columnwidth]{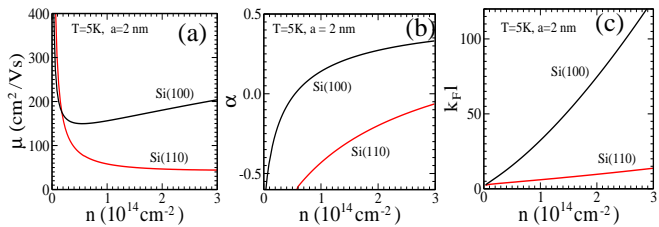}
	\caption{
(a) The calculated mobility in a Si:P $\delta$-doped layer as a
          function of density at a finite 
          temperatures, $T=5K$. 
The black (red) lines represent the mobility calculated with
parameters of Si(100) [Si(110)]. The well width $a=2$ nm is used in
this calculation. (b) The exponent of density 
dependent mobility calculated in (a), i.e. exponent in the
relation of $\mu \propto n^{\alpha}$. (c) $k_Fl$ as a function of
density where the mean free path is given by $l=v_F \tau$. 
	\label{fig:fig2}}
\end{figure}

In Fig.~\ref{fig:fig1} we show the 
calculated mobility of Si:P $\delta$-doped layers as a function of
density for different quantum 
well width $a=2$, 5, and 10 nm. For comparison we also show the results
(black lines)
calculated with Howard-Fang (HF) variational wave
function\cite{ando1982}. In Fig.~\ref{fig:fig1} 
charged impurities (dopants) are located at $z=0$ (i.e. the center of quantum well).
In the low density limit (or strong screening limit, $q_0=q_{TF}/2k_F \gg 1$) the
calculated mobility is inversely proportional to the density ($\mu \sim
1/n$) as the density 
increases upto $n_1 \sim g q_{TF}^2/16 \pi$. In this limit the effect of
the finite 2D layer width is small
since $k_Fa$ is small and the form factors are effectively unity.
Note that for Si(100) surface
orientation $q_{TF} \approx 0.13$ \AA$^{-1}$, which gives $n_1 \sim 1.3
\times 10^{13}$ cm$^{-2}$. For Si(110) surface orientation $q_{TF} \approx 0.52$
\AA$^{-1}$ and $n_1 \sim 4.2 \times 10^{14}$ cm$^{-2}$. When the
density increases further, $n>n_1$, mobility has a minimum value at $q_0 \sim
1$. In the weak screening high-density limit ($q_0 \ll 1$) 
the form factor of electron-impurity interaction arising from the finite width
of the 2D layer affects the density dependent
mobility since $k_F a>1$ now and the form factors become relevant.
For both $n>n_1$ and $n \gg  g/(16\pi a^2)$ the mobility
increases as $n^{1/2}$. In Fig.~\ref{fig:fig1} the non-monotonic
density dependent mobility as a function of density is
shown, i.e., as the carrier density increases mobility decreases first,
reached a minimum, and then it increases again. 
Thus, the three density regimes (low, intermediate, and high)
discussed analytically in the last section are clearly apparent in
Fig.~\ref{fig:fig1} with mobility decreasing first with increasing density, then
saturating around a minimum value around an intermediate density
before increasing again at higher density. 
The characteristic intermediate density for the mobility minimum
increases as the quantum well width decreases.
Fig.~\ref{fig:fig1} also shows the degeneracy dependent
mobility. At low densities $n < 10^{13}$ cm$^{-2}$ the mobility of
Si(110), which has a higher degeneracy factor and hence stronger screening,
is higher than that of Si(100), but at high densities the
mobility of Si(100) is higher. We can understand this degeneracy
dependence from the scattering time. From Eq.~(\ref{eq:taukf}) we
find\cite{hwang2012} $\mu \propto g^2$ for $q_0 
\gg1$ (strong screening limit) and $\mu \propto g^{-1}$ for $q_0 \ll
1$ (weak screening limit). 

In order to clearly see the high density behavior of mobility
we show in Fig.~\ref{fig:fig2}(a) the calculated mobility in linear scale  
as a function of density at a finite temperature $T=5K$.
The black (red) lines represent the mobility calculated with
parameters of Si(100) [Si(110)]. The well width $a=2$ nm is used in
this calculation and the impurities are located at
$z=0$. Fig.~\ref{fig:fig2}(b) shows the exponent of density  
dependent mobility for the results shown in (a), i.e. exponent in the
relation of $\mu \propto n^{\alpha}$. As described in Sec. II the
exponent $\alpha$ of Si(100) samples eventually approaches $1/2$ as the
carrier density 
increases. Fig.~\ref{fig:fig2}(c) shows the calculated $k_Fl$ as a
function of density where the  mean free path is defined by $l=v_F
\tau$. Note that $k_F l = (2\sigma/g) (h/e^2)$, where
the conductivity $\sigma$ has a unit $e^2/h$ and $g$ is the total
degeneracy. In terms of mobility we can express it as $k_F l = 0.83 \mu
\bar{n}/g$, where $\mu$ is measured in units of cm$^2/Vs$ and $\bar{n} =
n/(10^{14} cm^{-2})$. The calculated $k_F 
l$ increases super-linearly due to the finite width effect. 
In Sec. II we show that $\mu \propto n^{1/2}$ for $q_0 \ll 1$ and
$ k_Fa\gg1$. Thus in this high density limit we expect that the mean free
path increases linearly with density, $l \propto n$, and this behavior
is experimentally observed in Si:P $\delta$-doped layers \cite{goh2006}. However, 
in the strict 2D limit $k_F l$ increases linearly with density and
$l \propto \sqrt{n}$. The linear increase of $l$ with $n$ arises from
the finite width effect of the layer
as discussed analytically in Sec. II.




\begin{figure}
	\centering
	\includegraphics[width=1.\columnwidth]{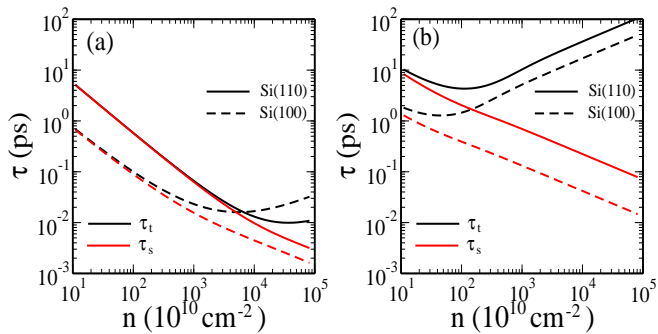}
	\caption{
(a) The transport scattering time (black lines), $\tau_t$, and single particle
relaxation time (red lines), $\tau_s$, as a function of
density for Si(100) (dashed lines) and Si(110) (solid lines) at $T=0$.
The charged impurities are located at $z=0$ (center of the well). The well width
$a=2$ nm is used in 
this calculation. In (b) we show the scattering times assuming the
impurities be outside the quantum well $z=5nm$.
	\label{fig:fig3}}
\end{figure}


So far we have only considered the transport scattering time ($\tau$) which
is relevant to the conductivity or mobility as shown in Sec. II. In
general, in the presence of disorder 
scattering there are two distinct relaxation
times\cite{dassarma1985}: the scattering lifetime or the transport relaxation time
($\tau$ or $\tau_t$)
and the quantum lifetime or the single particle relaxation time
($\tau_s$ or $\tau_q$). 
These two characteristic times of the system 
differ by the important $(1-\cos\theta)$ vertex correction factor. The
relaxation rate 
$\tau_s^{-1}$ is given by making the replacement $(1-\cos\theta)
\rightarrow 1$ in the integrand for the formula for $\tau^{-1}$ in
Eq.~(\ref{eq:tau_in}).
In general, $\tau_s$ determines the quantum level broadening, $\gamma
= \hbar/2\tau_s$, of the momentum eigenstates, and
physically, it simply represents the time between
scattering events between electron and impurity. 
Considering the difference between two scattering times we calculate
the single particle relaxation time and compare it with the transport
scattering time in following two figures.
We mention that just as the transport scattering time $\tau_t$ determines
the conductivity or the mobility of the system, the single particle
relaxation time $\tau_s$ determines level broadening as measured in
Shubnikov-de Haas (SdH) measurements --- in fact, the Dingle temperature
of SdH measurements is given by $\gamma$ defined above using $\tau_s$.  In
strong screening systems or for short-range disorder scattering (as in
3D metals or 2D Si MOSFETs), $\tau_s \sim \tau_t$ generally, and one does not
need to discuss two distinct impurity scattering times.  This is not
true in Si:P $\delta$-layers at high density where the system is weakly
screened, and the two scattering times could differ by a very large
factor in the presence of long-range Coulomb disorder.

\begin{figure}[t]
	\centering
	\includegraphics[width=\columnwidth]{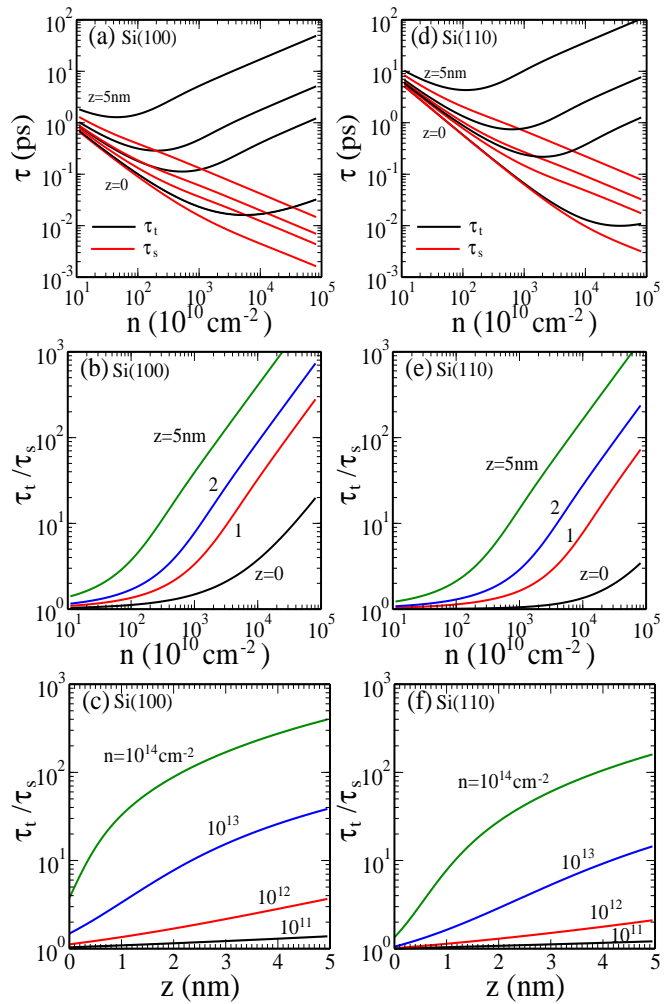}
	\caption{
(a), (b), and (c) show the calculated scattering times at $T=0$ with
parameters of Si(100). (d), (e), and (f) show the scattering times at
$T=0$ with
parameters of Si(110). The quantum well width $a=2$ nm is used in
this calculation. 
(a),(d) The transport scattering time (black lines), $\tau_t$, and
single particle 
relaxation time (red lines), $\tau_s$, as a function of density for different
impurity locations, $z=0$, 1, 2, 5nm (bottom to top).
(b),(e) The ratio $\tau_t/\tau_s$ as a function of
density for different impurity locations.
(c),(f) The ratio $\tau_t/\tau_s$ as a function of impurity location $z$
for different densities.
	\label{fig:fig4}}
\end{figure}

In Fig.~\ref{fig:fig3}(a) we show the transport scattering time,
$\tau=\tau_t$, and single particle relaxation time, $\tau_s$, of Si:P
$\delta$-doped layers as a function of
density for both Si(100) and Si(110).
The charged impurities are located at $z=0$ (center of the well). The
well width $a=2$ nm is used in 
this calculation. In Fig.~\ref{fig:fig3}(b) we show the scattering
times assuming the impurities to be outside the quantum well,
i.e. impurities at $z=5nm$. By replacing $(1-\cos\theta)$ with 1 in
Eq.~(\ref{eq:tau_in}) we have $\tau_s \approx 2 \tau_0 q_0^2/\pi \propto
n^{-1}$ for 
$q_0 \gg1$ (strong screening limit or low density limit).
In the opposite limit ($q_0 \ll 1$) we have $\tau_s \approx \tau_0 q_0
\propto n^{-1/2}$, assuming $n_i=n$. Thus, unlike $\tau$, as the carrier density
increases the calculated single
particle relaxation time $\tau_s$ decreases continuously without any
upturn which appears in the transport scattering time $\tau$ [see
Fig.~\ref{fig:fig3}(a)]. When the charged impurities are located
outside the quantum well (remote doping) the very strong enhancement
in the transport scattering time $\tau$ is found since large-angle scattering
($2k_F$ scattering) by the remote impurities (which is most weighted
in $\tau$) is strongly suppressed by the separation. 
The single particle relaxation time is also enhanced, but the
enhancement is much smaller than that of $\tau$ since all angles
contribute to $\tau_s$ [see Fig.~\ref{fig:fig3}(b)].

In Fig.~\ref{fig:fig4} we show the calculated scattering times and the
ration of $\tau$ to $\tau_s$ for both Si(100) and Si(110).
As shown in Figs.~\ref{fig:fig4}(a) and (d) the transport scattering
time, $\tau$, shows nonmonotonic behavior and strongly depends on the
location of dopant impurities. However, the single particle 
relaxation time, $\tau_s$, decreases monotonically as the density
increases, and has weaker dependence on impurity location.
In the strong screening limit and $z=0$ we have
$\tau = \tau_s \approx 2 \tau_0 q_0^2/pi \propto
n^{-1}$ and, therefore,$\tau/\tau_s \rightarrow 1$. At high densities
($q_0 \gg 1$) and $z=0$ we have $\tau/\tau_s \propto \sqrt{n}$, but
due to the factor $e^{-2k_F z}$ arising from the separation between
electrons and impurities the ratio is much enhanced as shown in
Fig.~\ref{fig:fig4}.


\begin{figure}[t]
	\centering
	\includegraphics[width=\columnwidth]{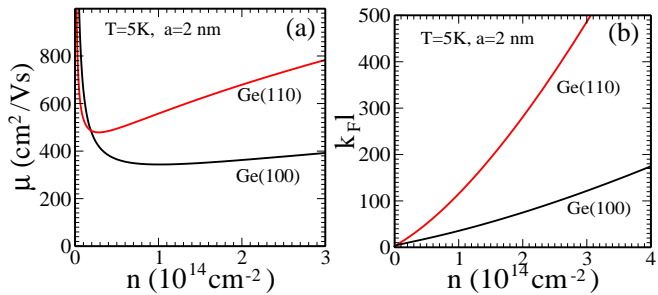}
	\caption{
(a) The calculated mobility in a Ge:P $\delta$-doped layer as a
function of density at a finite temperatures, $T=5K$. 
The black (red) lines represent the mobility calculated with
parameters of Ge(100) [Ge(110)]. The well width $a=2$ nm is used in
this calculation. (b) $k_Fl$ as a function of
density where the mean free path is given by $l=v_F \tau$. 
	\label{fig:fig5}}
\end{figure}

Finally, in Fig.~{\ref{fig:fig5}(a) we show the calculated 
mobility in a Ge:P $\delta$-doped layer as a
function of density at a finite temperature, $T=5K$. Recently
Ge:P $\delta$-doped layers have also attracted a lot of interest
due to the higher mobility in Ge than in Si \cite{scappucci2009}.
In this calculation we use the following parameters: effective mass
$m^*=0.12 m_e$, background dielectric constant $\kappa = 16$, and the
valley degeneracy $g_v=4$ for Ge(100) surface and $g_v=2$ for Ge(110)
surface. We use the well width $a=2$ nm and assume the dopants are
located at $z=0$. Fig.~\ref{fig:fig5}(b) shows $k_Fl$ as a function of
density where the mean free path is given by $l=v_F \tau$. 
We find that the mobility in a Ge:P $\delta$-doped layer is enhanced
by a factor of 10 compared with the mobility in a Si:P $\delta$-doped
layer. However, the overall density dependent transport behavior of the
Ge:P is very similar to that of Si:P. Due to the large value of the
effective Bohr radius in Ge ($a_B \sim 100$ \AA) the $q_0 =
q_{TF}/2k_F$ in Ge:P is much smaller than that of Si:P. Thus we find
that after reaching a minimum point in the density dependent mobility 
the upturn density is much smaller in a Ge:P $\delta$-doped layer
than that in a Si:P $\delta$-doped layer.

\section{discussion and conclusion}

Before concluding, we first want to discuss two issues not discussed
in the earlier sections with respect to our results.  These are the
important questions of comparison between experiment and our theory
and the issue of localization, which is of course experimentally
relevant, but is not explicitly included in our semiclassical
Boltzmann theory. 

First, in comparing our theory qualitatively to experiment, we note
that our theoretical results are at best an upper limit on the
experimental mobility since we have assumed the ionized dopants to be
the only source of resistive scattering whereas in reality there are
likely to be other sources of disorder (e.g. unintentional and hence
unknown impurities) in addition to the charged dopants.  We therefore
expect our results to agree with only the best available samples where
other sources of scattering are presumably suppressed.  This is indeed
the case.  The current highest measured mobility of Si:P (100)
$\delta$-layer comes from the Sandia National Lab\cite{bussmann}
where the samples
after annealing reached a mobility at 5K of close to 200 cm$^2$/Vs at a
carrier density of $n \sim 10^{14}$ cm$^{-2}$. According to our
Fig.~\ref{fig:fig1}, the 
mobility for the Si:P (100) $\delta$-layer varies between 200 and 500
cm$^2$/Vs which is consistent with the experimental data (and is in
fact in agreement with it if we use the lower value of the 2D
confinement width).  All other existing experimental mobility data in
the literature fall below our calculated value (but none above)
because the currently existing samples are not yet optimized and still
presumably contain substantial amount of unintentional quenched
impurities.  This conclusion is consistent with the Sandia finding
that the mobility increased substantially (by more than a factor of 2)
upon annealing which presumably got rid of some of the unintentional
impurities in the system.  In any case, our results presented in this
paper should motivate the experimentalists in producing density
dependent mobility data in Si and Ge $\delta$-doped 2D systems so that a
careful quantitative comparison can be carried out, which is not
possible at this stage because of the dearth of detailed experimental
data.

Second, we can discuss the onset of strong localization within our
Boltzmann theory by asking where the Ioffe-Reggel-Mott criterion\cite{lee1985} for
strong localization, $k_Fl=1$, starts to get satisfied in our system.
This will again provide us with an intrinsic limit on localization
since we assume $n=n_i$ which would underestimate the actual
localization density in real samples where $n<n_i$ is the likely
situation because of unintentional impurities.  We note that our
calculated $k_Fl$ [Fig.~\ref{fig:fig2}(c)] increases monotonically with density,
and thus the localization condition of $k_Fl$ is a low-density
condition, which is consistent with experimental data where the
localization behavior is seen only for $n<10^{13}$ cm$^{-2}$. In fact, for
$n=10^{14}$ cm$^{-2}$, we get $k_Fl \sim 40$ in Si:P (100) 2D system, which is far
from the strongly localized regime.  Using our simple formula $k_Fl
=0.83\mu \tilde{n}$ derived in Section III, we see from Fig.~\ref{fig:fig1}(a) that
for $n=10^{13}$ cm$^{-2}$ (where our calculated $\mu= 200$ cm$^2$/Vs),
$k_Fl \sim 4$
which is rather close to the strong localization condition.
Therefore, it is indeed possible that the actual localization in
realistic samples (where the mobility is likely to be lower than our
calculated results assuming that the ionized dopants are the only
scattering source) occurs around $n \sim 10^{13}$ cm$^{-2}$ carrier density.

We also comment on our finding of a mobility minimum at a
characteristic density $n_c$ in the 2D $\delta$-doped system.  Since $n_c$
defines the crossover from the strong screening to the weak screening
behavior in the system, we can crudely obtain $n_c$ from the condition
$q_0=1$, i.e. $q_{TF}=2k_F$.  This happens for Si(100) system at
$n_c=1.2 \times 10^{13}$ cm$^{-2}$, which is roughly where the mobility starts to go
from a decreasing function of density to an increasing function of
density in Fig.~\ref{fig:fig1}. It will be very interesting to observe this strong
to weak screening crossover directly experimentally.

In conclusion, we have studied the effect of doping density on
electronic transport in 2D Si:P $\delta$-doped layers.
Since the carrier density is equal to the dopant impurity density
we find that the density dependent transport is qualitatively different from
that of Si-MOSFETs where the carrier density is independent of the impurity
density since carriers are induced by a gate and the impurities are
unintentional oxide charges at the interface. The density dependent
mobility of a Si:P $\delta$-doped layer 
is also different from that of a modulation $\delta$-doped  2D GaAs
quantum well system
because of the physical separation between electrons and dopants in
the GaAs system. 
We find that the density dependent
mobility of the Si:P system shows a non-monotonic behavior. 
At $q_0 = q_{TF}/2k_F \sim 1$ the mobility has a minimum value.
However, the calculated mean free path increases monotonically with
increasing density as observed experimentally \cite{goh2006}.
We also calculate both the transport scattering time $\tau$ and
the single particle relaxation time $\tau_s$ in Si:P $\delta$-doped
layers. The $\tau$ shows nonmonotonic behavior like mobility and
strongly depends on the location of impurities, but the
$\tau_s$ keeps decreasing with increasing density and is less
sensitive to the location of impurity.

In this paper we have simplified the confinement potential as a square
quantum well and neglected two important features of high density Si:P samples; the
localization effects and the multisubband 
effects. As shown in Fig.~\ref{fig:fig1} our results are not sensitive to
the shape of the confinement potential, and given that the exact
impurity and dopant configurations in real samples are not known
accurately, we do not see much point in trying to improve our
confinement model. 
Since the localization effects are observed at very low densities ($n
\alt 10^{13}$ cm$^{-2}$) our results 
may not be applicable to the actual experiment in this density range.
Usually in Si:P $\delta$-doped layers only a few subbands are filled.
Even though the mobility of carriers depends on the intersubband
scattering \cite{ando1982} we expect that our results quantitatively explain
the observed experimental data.
In particular, as emphasized throughout this paper, our work provides
the upper limit on the possible achievable mobility in delta-doped 2D
systems since we include the only source of scattering that must
always be present in the system, namely, the scattering by the ionized
dopants contributing the electrons to the system.  The experimental
mobility can be lower than our calculated results, but it can never be
higher.  Our estimate of the localization density using the
Ioffe-Regge-Mott criterion indicates that high-quality intrinsic
delta-doped samples with very little unintentional background
impurities may very well show effective metallic behavior at carrier
density $10^{13}$ cm$^{-2}$ or lower, but much of the strongly metallic
behavior most likely manifests for density around $10^{14}$ cm$^{-2}$ or
higher.  Our work should motivate detailed experimental transport
studies for this interesting and novel class of 2D electron systems
which allow the study of 2D transport at very high carrier density
along with very high impurity density, a transport regime little
studied before in 2D electron systems.

\section*{acknowledgments}
This work was supported by LPS-NSA.

\end{document}